\documentclass[a4paper,11pt]{article}
\pdfoutput=1 

\usepackage{jcappub} 

\usepackage[T1]{fontenc} 

\usepackage{graphicx}
\usepackage{xcolor}
\usepackage{tikz}
\usepackage[caption=false]{subfig}
\usepackage{mathrsfs,mathtools}
\usepackage{physics,amssymb}
\usepackage{bm}
\usepackage{braket}
\usepackage{listings}
\usepackage{cases}
\usepackage{comment}
\usepackage{soul}
\usepackage{cancel}
\usepackage{cases}
\usepackage[utf8]{inputenc}
\usepackage{url}
\usepackage{longtable}
\usepackage[normalem]{ulem}
\usepackage{xspace}
\usepackage{acronym}

\newcommand{\ee}{\mathrm{e}}
\newcommand{\diag}{\mathrm{diag}}

\newcommand{\Mpl}{M_\mathrm{Pl}}
\newcommand{\ns}{n_{\mathrm{s}}}

\newcommand{\LC}[3]{\qty\Big{\smqty{#1 \\ #2 #3}}}

\newcommand{\sur}{\mathrm{sur}}
\newcommand{\uend}{\mathrm{end}}

\newcommand{\uJ}{\mathrm{J}}

\newcommand{\calL}{\mathcal{L}}

\newcommand{\calM}{\mathcal{M}}

\newcommand{\calO}{\mathcal{O}}

\newcommand{\calP}{\mathcal{P}}
\newcommand{\calR}{\mathcal{R}}

\newcommand{\Mg}{M_g}
\newcommand{\MG}{M_\Gamma}

\newcommand{\bae}[1]{\begin{align} #1 \end{align}}

\newcommand{\beae}[1]{\begin{equation}\begin{aligned} #1 \end{aligned}\end{equation}}

\newcommand{\bme}[1]{\begin{multline} #1 \end{multline}}
\newcommand{\bmte}[1]{\begin{multlined}[t] #1 \end{multlined}}

\definecolor{MONZA}{HTML}{CF000F}
\definecolor{DARKBLUE}{HTML}{00008b}
\definecolor{DARKMAGENTA}{HTML}{8b008b}

\acrodef{EoM}{equation of motion}
\acrodef{CMB}{cosmic microwave background}
\acrodef{EFT}{effective field theory}
\acrodef{SM}{Standard Model}
\acrodef{UV}{ultraviolet}

\title{Hybrid metric-Palatini Higgs inflation}

\author[a]{Minxi He,}
\author[b]{Yusuke Mikura,}
\author[c,b,a]{and Yuichiro Tada}

\affiliation[a]{Theory Center, IPNS, KEK, 1-1 Oho, Tsukuba, Ibaraki 305-0801, Japan}
\affiliation[b]{Department of Physics, Nagoya University,\\ Furo-cho Chikusa-ku, Nagoya 464-8602, Japan}
\affiliation[c]{Institute for Advanced Research, Nagoya University,\\ Furo-cho Chikusa-ku, Nagoya 464-8601, Japan}

\emailAdd{heminxi@post.kek.jp}
\emailAdd{mikura.yusuke.s8@s.mail.nagoya-u.ac.jp}
\emailAdd{tada.yuichiro.y8@f.mail.nagoya-u.ac.jp}

\abstract{
We propose an extension of the Higgs inflation to the hybrid metric-Palatini gravity, where we introduce non-minimal couplings between Higgs and both the metric-type and the Palatini-type Ricci scalars. We study the inflationary phenomenology of our model and find that slow-roll inflation can be realized in the large-field regime, giving the observationally favored predictions. In particular, the scalar spectral index exhibits an attractor behavior to $\ns\sim 0.964$, while the tensor-to-scalar ratio can take an arbitrary value depending on the non-minimal coupling parameters, with the metric-Higgs limit $r\sim10^{-3}$ being the maximum. We also investigate the unitarity property of our model. As the \ac{UV} cutoff as a low-energy \ac{EFT} of this model is significantly lower than the Planck scale due to a strong curvature of field-space, we consider a possible candidate of UV-extended theories with an additional scalar field introduced so as to flatten the field-space in five-dimension. While the field-space can be flatten completely and this approach can lead to a weakly-coupled \ac{EFT}, we gain an implication that Planck-scale \ac{EFT} can be only realized in the limit of metric-Higgs inflation. We also discuss generalizations of the model up to mass-dimension four.
}

\begin{document} 

\begin{flushright}
    KEK-TH-2451 \\ 
    KEK-Cosmo-0297
\end{flushright}

\maketitle
\flushbottom
\acresetall
\section{Introduction}
Nowadays cosmic inflation is recognized as a part of standard cosmology, not only giving natural solutions to several major problems of the Big-Bang cosmology but also seeding the large-scale structure observed today. Since the advent of its concept, a huge number of theoretical investigations have been made so far to identify the origin of inflation with various motivations. Among the plethora of proposals, the Higgs inflation~\cite{Bezrukov:2007ep,Barvinsky:2008ia} has been attracting a lot of attention because in this scenario the inflaton(s) is identified with the Higgs fields which are the only fundamental scalars in the \ac{SM} of particle physics. In the Higgs inflation, a large non-minimal coupling between the Higgs fields and gravity is introduced in addition to the usual Higgs sector. This coupling plays a significant role in realizing a flat potential that is suitable for a successful inflation scenario.

The Higgs inflation has several variants because the gravitational sector can be interpreted in several ways. A commonly-used interpretation is called the metric formalism in which the underlying geometry is the (pseudo)-Riemannian and two assumptions, metric-compatibility and torsion-free, are afforded to the connection $\Gamma$ by hand. Given these two assumptions, the affine connection is uniquely determined as the Levi-Civita connection defined by
\bae{
\LC{\lambda}{\mu}{\nu}=\frac{1}{2}g^{\lambda\sigma}\left(\partial_\mu g_{\sigma\nu}+\partial_\nu g_{\mu\sigma}-\partial_\sigma g_{\mu\nu}\right).
}
Geometrical objects such as the Ricci tensor $R_{\mu\nu}[g]$ and curvature $R[g]=g^{\mu\nu}R_{\mu\nu}[g]$ are therefore defined as functions of the metric. From a theoretical standpoint, however, one does not necessarily adopt the metric formalism but can interpret gravity in, e.g., the Palatini formalism~\cite{Einstein:1925:EFG,Palatini}. In this formalism, the connection is not given a priori by the Levi-Civita one, but rather determined by the Euler--Lagrange equation. Contrary to the metric formalism, the Ricci tensor $R_{\mu\nu}[\Gamma]$ is defined as a function solely of the connection and accordingly the definition of the Ricci scalar is altered as $R[\Gamma]=g^{\mu\nu}R_{\mu\nu}[\Gamma]$ (note that we denote the Ricci scalar formed in the Palatini way as $R[\Gamma]$ for simplicity while the Ricci scalar also depends on the metric). If the gravity sector is merely given by the Einstein--Hilbert term, $\Mpl^2R[g]$ in the metric formalism or $\Mpl^2R[\Gamma]$ in the Palatini one, these two formalisms are equivalent because the general affine connection is determined to be the Levi-Civita one by the Euler--Lagrange equation up to an arbitrary vector.\footnote{More precisely, the arbitrary vector comes from the projective symmetry of the Ricci scalar, namely $ R \to R$ under the projective transformation $ \Gamma^{\lambda}_{\mu\nu} \to \Gamma^{\lambda}_{\mu\nu} + \delta^{\lambda}_{\nu} U_{\mu} $. This vector can be understood as a gauge degree of freedom that can be killed by the gauge fixing. In this paper, we eliminate the gauge degree by imposing the torsion-less condition.} However, when the theory of interest contains a non-minimal coupling between matter field(s) and the affine connection as in the Higgs inflation, two formalisms lead to different physical consequences even though apparent forms of the Lagrangian are the same. Corresponding to the choice of formalisms, metric or Palatini, there are two variants of the Higgs inflation: metric-Higgs and Palatini-Higgs inflation~\cite{Bauer:2008zj}.

Recently, the metric-Higgs and Palatini-Higgs inflation have been studied focusing more on their theoretical aspects in the high energy regime. Phenomenology of both models has been studied in depth and it is known that inflationary predictions of either are perfectly consistent with current observations of the \ac{CMB} (see Refs.~\cite{Rubio:2018ogq,Tenkanen:2020dge} for recent reviews). To fit the observations, however, the non-minimal coupling constant between the Higgs fields and gravity should be quite large unless the self-coupling of the Higgs is tuned to be extremely small. This large coupling can be problematic from a viewpoint of theoretical consistency because it can result in a very low \ac{UV} cutoff above which perturbation theory is no longer valid and may induce a violation of perturbative unitarity~\cite{Burgess:2009ea,Barbon:2009ya,Barvinsky:2009ii,Burgess:2010zq,Hertzberg:2010dc,Bezrukov:2010jz,Bauer:2010jg,Lerner:2011it,Kehagias:2013mya,Ren:2014sya,McDonald:2020lpz,Hamada:2020kuy,Antoniadis:2021axu}. Indeed, in the metric-Higgs inflation, longitudinal modes of the gauge bosons acquire a spiky effective mass during preheating which leads to the production of gauge bosons with energy higher than the cutoff scale~\cite{Ema:2016dny}. Whereas, in the case of the Palatini-Higgs scenario, the unitarity issue would not be observed during preheating even though the UV cutoff is considerably smaller than the Planck scale~\cite{Rubio:2019ypq}. It validates that the whole dynamics can be described by the given low-energy \ac{EFT}. While there might be no practical problem for phenomenology, it is of theoretical interest to investigate its \ac{UV} aspects as the low \ac{UV} cutoff implies that a new physics would intervene at the scale.

Given the situation, \ac{UV}-completion is an interesting topic of study in both models. Understanding of \ac{UV}-completion for the metric-Higgs inflation has been developed in recent years~\cite{Lerner:2010mq,Giudice:2010ka,Barbon:2015fla,Lee:2018esk,Ema:2020zvg}. In particular, it was found in Refs.~\cite{Giudice:2010ka,Ema:2020zvg} that the metric-Higgs inflation can be viewed as a non-linear sigma model and the low \ac{UV} cutoff originates in the strong curvature of field-space spanned by the Higgs fields. A corresponding UV-completing sigma field is identified with a new scalar degree of freedom arising from an additional $R^2[g]$ operator that is required for one-loop renormalizability~\cite{Salvio:2015kka,Ema:2017rqn,Gorbunov:2018llf}. It lifts up the UV cutoff to the Planck scale by flattening the curved field-space and also diminishes the spiky effective mass of longitudinal modes of gauge fields, removing the unitarity issue during the preheating process~\cite{He:2018mgb}. On the other hand, this cannot be simply applied to the Palatini case because the quadratic operator $R^2[\Gamma]$ in the Palatini formalism does not yield a dynamical degree of freedom~\cite{Edery:2019txq} (See also Refs.~\cite{Meng:2004yf,Antoniadis:2018ywb,Tenkanen:2019jiq,Tenkanen:2020xbb,Gialamas:2019nly,Gialamas:2020snr,Lykkas:2021vax,Gialamas:2021enw} for relevant works). Furthermore, two of the authors previously showed by taking a similar approach of Ref.~\cite{Giudice:2010ka} that the cutoff scale is still low even though a new scalar field (added by hand) eliminates the curvature of the field-space~\cite{Mikura:2021clt}. As a possible \ac{UV}-completion is yet to be specified, further investigations are required for satisfactory understanding of the Palatini-Higgs inflation.\footnote{See e.g. Ref.~\cite{Aoki:2022csb} for an interesting direction of research.}

A generalization of the Palatini-Higgs inflation may give us insights into the difference on UV-behaviors. The Palatini-Higgs inflation is a minimal theory in that the action only possesses the Palatini curvature $R[\Gamma]$ in a linear form. Nevertheless, there are no fundamental reasons to exclude the metric curvature $R[g]$ in the action. Gravity theories with both curvatures are described in a framework of the hybrid metric-Palatini gravity~\cite{Harko:2011nh} which has been applied mainly to the higher-curvature theories known as $f(R)$ theories (see Ref.~\cite{Capozziello:2015lza} for a review). In this paper, we apply the generalized formalism to the Higgs inflation and propose a new inflationary scenario which bridges the metric-Higgs and the Palatini-Higgs inflation continuously. We call the generalization \emph{hybrid metric-Palatini Higgs inflation}. Here we note that a relevant work on inflationary phenomenology in the hybrid metric-Palatini framework~\cite{Asfour:2022qap} independently comes out simultaneously with this paper.

This paper is organized as follows. In Sec.~\ref{Hybrid metric-Palatini Higgs inflation}, we first define the hybrid metric-Palatini Higgs inflation in the Jordan frame and derive an equivalent metric theory in the Einstein frame. We next study its inflationary phenomenology and discuss the theoretical consistency. A \ac{UV}-extension of the model is investigated in Sec.~\ref{Sec. UV-completion}. We summarize the embedding approach as a way of \ac{UV}-extension and apply the procedure to our model. We discuss possible generalizations of our model in Sec.~\ref{Sec. Generalizatons}. The conclusions are presented in Sec.~\ref{Sec. Conclusions}. Throughout the paper we adopt the natural unit $c=\hbar=1$ and $\eta_{\mu\nu}=\diag(-1,1,1,1)$ is used as the sign of the Minkowski metric.

\section{Hybrid metric-Palatini Higgs inflation}\label{Hybrid metric-Palatini Higgs inflation}
We define the hybrid metric-Palatini Higgs inflation in the Jordan frame as the following Lagrangian (with a subscript $\uJ$) 
\bae{\label{Eq. action for the hybrid metric-Palatini}
\frac{\calL}{\sqrt{-g_{\uJ}}}=\frac{\Mpl^2}{2}\left(1+\xi_\Gamma\frac{\phi_{\uJ i}^2}{\Mpl^2}\right)R_\uJ[\Gamma]+\frac{1}{2}\xi_g \phi_{\uJ i}^2R_\uJ[g]-\frac{1}{2}g^{\mu\nu}_{\uJ}\partial_\mu\phi_{\uJ i}\partial_\nu\phi_{\uJ i}-\frac{\lambda}{4}(\phi_{\uJ i}^2)^2,
}
where $\Mpl=1/\sqrt{8\pi G}$ is the reduced Planck mass, $\phi_{\uJ i}$ ($i=1, \cdots, 4$) represent four real degrees of freedom of the Higgs fields, $\lambda$ is their self-coupling, $R_\uJ[\Gamma]$ and $R_\uJ[g]$ are Ricci scalars in the Palatini and the metric formalisms respectively, and $\xi_\Gamma$ and $\xi_g$ are positive non-minimal coupling constants between Higgs and the two different Ricci scalars. Note that $R_\uJ[\Gamma]$ is defined as 
\bae{
    R_\uJ[\Gamma] \coloneqq g_{\uJ}^{\mu\nu} R_{{\uJ}\mu\nu} [\Gamma],
}
where $\Gamma$ is a general affine connection determined via the Euler--Lagrange equation, while the affine connection is replaced by the Levi-Civita connection in $R_\uJ[g]$. Here it should be stressed again that the Palatini curvature is not only a function of the connection but it also depends on the metric, so that the expression $R_\uJ[\Gamma]$ is just for the sake of brevity. In the action, we neglect the mass term of the Higgs fields because it is irrelevant to our discussion.

This action encompasses the metric-Higgs and Palatini-Higgs inflation scenarios, each of which can be obtained in the limit of $\xi_\Gamma=0$ or $\xi_g=0$. When $\xi_\Gamma=0$, $R_\uJ[\Gamma]$ is minimally coupled to the Higgs fields such that $\Gamma$ is eventually equivalent to the Levi-Civita one according to the Euler--Lagrange constraint. Thus, $R_\uJ[\Gamma]$ can be replaced by $R_\uJ[g]$, giving the metric-Higgs scenario~\cite{Bezrukov:2007ep}. When $\xi_g=0$, the action is reduced to the usual definition of the Palatini-Higgs inflation~\cite{Bauer:2008zj}.

Given the action~\eqref{Eq. action for the hybrid metric-Palatini}, one can see that the connection $\Gamma$ is not a dynamical degree of freedom because the action only involves its first derivatives. Therefore, with use of the constraint equation for $\Gamma$, we can obtain an equivalent theory of Eq.~\eqref{Eq. action for the hybrid metric-Palatini} in the metric formalism, which is convenient for the study of its phenomenology and \ac{UV} aspects. By varying the action with respect to $\Gamma$, the constraint equation reads
\bae{\label{Eq. EoM for connection} 
\overset{\Gamma}{\nabla}_\lambda\left[F\sqrt{-g_\uJ}g_\uJ^{\mu\nu}\right]=0,
}
where $\overset{\Gamma}{\nabla}$ is the covariant derivative associated with the general connection $\Gamma$ and $F\coloneqq1+\xi_\Gamma\frac{\phi_{\uJ i}^2}{\Mpl^2}$. Assuming the torsion-free condition, it can be algebraically solved and its constraint equation is given by
\bae{\label{Eq. sol of Gamma}
\Gamma^{\lambda}_{\ \mu\nu}=\LC{\lambda}{\mu}{\nu}+\frac{1}{2}\left[\delta^\lambda_\mu\partial_\nu\log F+\delta^\lambda_\nu \partial_\mu \log F -g_{\uJ\mu\nu}g_\uJ^{\lambda\sigma}\partial_\sigma\log F\right],
}
where the first term on the right hand side denotes the Levi-Civita connection defined by $g_{\uJ\mu\nu}$. The constraint equation relates the Palatini curvature $R_\uJ[\Gamma]$ with the metric one $R_\uJ[g]$ as
\bae{
R_\uJ[\Gamma]=R_\uJ[g]+\frac{6\xi_\Gamma^2}{\left(1+\xi_\Gamma\frac{\phi_{\uJ i}^2}{\Mpl^2}\right)^2}\frac{\phi_{\uJ i}\phi_{\uJ j}}{\Mpl^4}\partial_\mu \phi_{\uJ i}\partial^\mu \phi_{\uJ j},
}
and it gives an equivalent metric theory of Eq.~\eqref{Eq. action for the hybrid metric-Palatini} in the Jordan frame as
\bae{
\frac{\calL}{\sqrt{-g_{\uJ}}}=\frac{\Mpl^2}{2}\left[1+(\xi_\Gamma+\xi_g)\frac{\phi_{\uJ i}^2}{\Mpl^2}\right]R_\uJ[g]-\frac{1}{2}\left(\delta_{ij}-\frac{6\xi_\Gamma^2\frac{\phi_{\uJ i}\phi_{\uJ j}}{\Mpl^2}}{1+\xi_\Gamma\frac{\phi_{\uJ i}^2}{\Mpl^2}}\right)g_\uJ^{\mu\nu}\partial_\mu \phi_{\uJ i}\partial_\nu \phi_{\uJ j}-\frac{\lambda}{4}(\phi_{\uJ i}^2)^2.
}
Hereafter, we write $R_\uJ[g]$ as simply $R_\uJ$ in the metric theory unless necessary.

The non-minimal coupling can be removed from the theory through the Weyl transformation defined by
\bae{
g_{\mu\nu}\coloneqq \Omega^2 g_{\uJ\mu\nu}=\left[1+(\xi_\Gamma+\xi_g)\frac{\phi_{\uJ i}^2}{\Mpl^2}\right]g_{\uJ\mu\nu}.
}
Accompanied by the metric rescaling, the Ricci curvature non-trivially transforms as
\bae{
\sqrt{-g_\uJ}\Omega^2 R_\uJ=\sqrt{-g}\left(R-6\Omega g^{\mu\nu}\nabla_\mu\nabla_\nu\Omega^{-1}\right),
}
and hence the resulting Einstein-frame expression is given by
\bae{\label{Eq. E-frame action}
\frac{\calL}{\sqrt{-g}}=\frac{\Mpl^2}{2}R-\frac{1}{2}G_{ij}g^{\mu\nu}\partial_\mu\phi_{\uJ i}\partial_\nu\phi_{\uJ j}-\frac{\lambda (\phi_{\uJ i}^2)^2}{4\left\{1+(\xi_\Gamma+\xi_g)\frac{\phi_{\uJ k}^2}{\Mpl^2}\right\}^2},
}
where $G_{ij}$ denotes the field-space metric whose explicit form is given by
\bae{\label{Eq. HI field-space}
G_{ij}\coloneqq \frac{\delta_{ij}}{1+(\xi_\Gamma+\xi_g)\frac{\phi_{\uJ k}^2}{\Mpl^2}}+\frac{6\xi_g\left\{2\xi_\Gamma+\xi_g+\xi_\Gamma(\xi_\Gamma+\xi_g)\frac{\phi_{\uJ k}^2}{\Mpl^2}\right\}}{\left(1+\xi_\Gamma\frac{\phi_{\uJ k}^2}{\Mpl^2}\right)\left\{1+(\xi_\Gamma+\xi_g)\frac{\phi_{\uJ k}^2}{\Mpl^2}\right\}^2}\frac{\phi_{\uJ i}\phi_{\uJ j}}{\Mpl^2}.
}

\subsection{Inflationary phenomenology}\label{sec: inflationary phenomenology}
In this section we investigate the inflationary phenomenology of our model. To this end, we take the unitary gauge for the Higgs fields and neglect gauge interactions, giving the following Einstein-frame Lagrangian for the radial mode $\phi_\uJ$:
\bae{\label{Eq. Inf. Pheno. lagrangian}
\frac{\calL}{\sqrt{-g}}=\frac{\Mpl^2}{2}R-\frac{1}{2}F^2(\phi_\uJ)g^{\mu\nu}\partial_\mu\phi_\uJ\partial_\nu\phi_\uJ-\frac{\lambda\phi_\uJ^4}{4\left\{1+(\xi_g+\xi_\Gamma)\frac{\phi_\uJ^2}{\Mpl^2}\right\}^2},
}
with the radial eigenvalue $F^2(\phi_\uJ)$ of the field-space metric:
\bae{
F^2(\phi_\uJ)=\frac{1+(\xi_g+2\xi_\Gamma+6\xi_g^2+12\xi_g\xi_\Gamma)\frac{\phi_\uJ^2}{\Mpl^2}+\left(\xi_g\xi_\Gamma+\xi_\Gamma^2+6\xi_g^2\xi_\Gamma+6\xi_g\xi_\Gamma^2\right)\frac{\phi_\uJ^4}{\Mpl^4}}{\left(1+\xi_\Gamma\frac{\phi_\uJ^2}{\Mpl^2}\right)\left\{1+(\xi_g+\xi_\Gamma)\frac{\phi_\uJ^2}{\Mpl^2}\right\}^2}.
}
The non-trivial kinetic term can be made canonical by redefining the scalar field through
\bae{\label{Eq. canonical normalization}
    \dv{\phi_\uJ}{\chi}\coloneqq \frac{1}{F}.
}
Generally speaking, an analytic expression of $\chi$ can hardly be obtained by doing the integration. However, an asymptotic form of $F^2(\phi_\uJ)$ in the large field limit $\phi_\uJ\gg\Mpl$ is still useful to analyze the dynamics. In the large field limit, the function $F^2$ can be approximated as 
\bae{\label{eq: K(phiJ) in the large field limit}
    F^2(\phi_\uJ)\simeq\frac{1+6\xi_g}{\xi_g+\xi_\Gamma}\frac{\Mpl^2}{\phi_\uJ^2}.
}
Combining with Eq.~\eqref{Eq. canonical normalization}, the radial mode $\phi_\uJ$ is related with the new field $\chi$ as
\bae{
\frac{\phi_\uJ}{\Mpl}\simeq Z\ee^{\sqrt{\frac{\xi_g+\xi_\Gamma}{1+6\xi_g}}\frac{\chi}{\Mpl}},
}
where $Z$ is an integration constant. In terms of $\chi$, the potential $U$ is written as
\bae{\label{Eq. pheno potential}
    U\simeq \frac{\lambda \Mpl^4}{4(\xi_g+\xi_\Gamma)^2}\left\{1-\frac{2}{Z^2(\xi_g+\xi_\Gamma)}\ee^{-2\sqrt{\frac{\xi_g+\xi_\Gamma}{1+6\xi_g}}\frac{\chi}{\Mpl}}\right\}.
}
Note that the constant $Z$ can not be fixed because the function $F^2(\phi_\uJ)$ was truncated for the large field dynamics. However the undetermined constant does not appear in observables because it can be absorbed into the number of e-folds $N$ as we shall see below.

Inflationary observables are often characterized by potential slow-roll parameters
\bae{
\epsilon\coloneqq\frac{\Mpl^2}{2}\left(\frac{1}{U}\dv{U}{\chi}\right)^2,\qquad \eta\coloneqq\Mpl^2\left(\frac{1}{U}\dv[2]{U}{\chi}\right),
}
and the number of backward e-folds $N$ defined by
\bae{
N(\chi)\coloneqq \frac{1}{\Mpl^2}\int^\chi_{\chi_\mathrm{end}} U\left(\dv{U}{\chi}\right)^{-1}\dd{\chi}.
}
Here $\chi_\mathrm{end}$ denotes the field value of $\chi$ at the end of inflation. The primary cosmological observables are the spectral index $\ns$ and the tensor-to-scalar ratio $r$, that are expressed to the leading order in the potential slow-roll parameters as
\bae{
    \ns\simeq 1-6\epsilon+2\eta, \qquad r\simeq 16\epsilon.
}
By requiring that inflation lasts for a sufficiently long period, the field value at the end of inflation becomes negligible and the backward e-folds $N$ can be well approximated as
\bae{
N\simeq \frac{Z^2(1+6\xi_g)}{8}\ee^{2\sqrt{\frac{\xi_g+\xi_\Gamma}{1+6\xi_g}}\frac{\chi}{\Mpl}}.
}
With use of the number of e-folds $N$, the potential slow-roll parameters can be written without depending on the constant $Z$ as
\bae{
    \epsilon\simeq \frac{1+\xi_g}{8(\xi_\Gamma+\xi_g)N^2}, \qquad \eta\simeq -\frac{1}{N},
}
and then the resulting predictions are given by
\bae{\label{Eq: observables}
    \ns\simeq 1-\frac{2}{N},\qquad r\simeq\frac{1+6\xi_g}{\xi_g+\xi_\Gamma}\frac{2}{N^2},
}
where we assume $\epsilon\ll\abs{\eta}$ for sufficiently large $N$ in the expression of the spectral index. Regardless of parameter choices, the spectral index $\ns$ exhibits an attractor behavior converging to $\ns\simeq 0.964$ (for a typical value $N\simeq 55$) which is well consistent with the \ac{CMB} observation by the Planck collaboration~\cite{Planck:2018jri}. On the other hand, the tensor-to-scalar ratio behaves differently, particularly with a large $\xi_\Gamma$. In the limit of $\xi_\Gamma=0$ corresponding to the metric-Higgs scenario, the tensor-to-scalar ratio also converges to a constant of order $\calO(10^{-3})$ which can be tested by future \ac{CMB} observations~\cite{CMB-S4:2016ple,Hazumi:2019lys}. Whereas, in the Palatini-Higgs limit $\xi_g=0$, as it is suppressed linearly by the large $\xi_\Gamma$, the power spectrum of the primordial gravitational waves can be extremely small.
We have confirmed the above $N$-dependence of slow-roll parameters by full numerical calculations of the background dynamics.

Note that the two non-minimal coupling parameters cannot be arbitrary chosen because they are constrained by the observation on the dimensionless power spectrum of curvature perturbation $\calP_\zeta$:
\bae{\label{Eq: curvature perturbation}
\calP_\zeta=\frac{1}{24\pi^2\Mpl^4}\frac{U}{\epsilon}\simeq \frac{1}{(1+6\xi_g)(\xi_g+\xi_\Gamma)}\frac{\lambda N^2}{12\pi^2}.
}
An observed amplitude of the power spectrum $\calP_\zeta\sim2.1\times 10^{-9}$~\cite{Planck:2018vyg,Planck:2018jri} yields a relation between $\xi_\Gamma$ and $\xi_g$, eliminating one parameter in the tensor-to-scalar ratio (with fixed $\lambda$).

\subsection{Perturbative Unitarity}
The inflationary phenomenology investigated in the previous subsection is based on the assumption that the theory~\eqref{Eq. Inf. Pheno. lagrangian} is under the \ac{EFT} control during inflation and the succeeding reheating phase. However, it has been pointed out in previous works~\cite{Ema:2016dny,Hamada:2020kuy} that the large non-minimal coupling between the Higgs fields and the Ricci curvature may induce a (perturbative) unitarity violation due to gauge bosons excited during preheating.\footnote{There is an exception called the critical Higgs scenario~\cite{Hamada:2014iga,Bezrukov:2014bra,Hamada:2014wna,Enckell:2020lvn}. In this scenario, the Higgs self-coupling is tuned to be small along with the renormalization group and the non-minimal coupling can be small, which makes the theory free from the unitarity problem.} If the unitarity violation occurs, the perturbative analysis becomes no longer reliable and thus we need to specify a more fundamental theory to have precise predictions. In this subsection, we compare the UV cutoff scales and physical energy scales explicitly, and show that our model also requires \ac{UV}-extension.

\subsubsection{Cutoff scales of the theory}
Recall that the Lagrangian in the Einstein frame is given by Eq.~\eqref{Eq. E-frame action}. In the Einstein frame, the unitarity violation scales can be estimated from geometry of the field-space $\calM_4$ spanned by the four scalar Higgs fields because the scattering amplitude is associated with geometrical objects of the field-space $\calM_4$~\cite{Alonso:2015fsp,Nagai:2019tgi,Mikura:2021clt}.\footnote{One can instead work in the Jordan frame, in which case the cutoff scales can be derived by explicitly calculating the scattering between the graviton and the Higgs fields. See, e.g., Refs.~\cite{Antoniadis:2021axu,Ito:2021ssc,Karananas:2022byw} for the detail. According to Ref.~\cite{Ema:2020zvg}, a better way to discuss the field-space is to use the conformal frame where the conformal mode of the metric decouples with other scalar fields. All the scalar fields including the conformal mode are conformally coupled with the tensor modes in this frame. However, we expect that the curvature of the field-space in the Einstein frame will be very similar to the one in the conformal frame in our consideration because the only difference between minimal coupling and conformal coupling is a factor $1/6$ which will not change the curvature significantly.} We determine the \ac{UV} cutoff $\Lambda$ by the Ricci curvature through $\Lambda\sim |\calR_G|^{-1/2}$. Given the field-space metric~\eqref{Eq. HI field-space}, the field-space curvature $\calR_G$ can be written as
\bae{\label{eq: RG}
    \calR_G=\frac{c_1+c_2 \phi_{\uJ k}^2/\Mpl^2+c_3 \phi_{\uJ k}^4/\Mpl^4+c_4 \phi_{\uJ k}^6/\Mpl^6+c_5 \phi_{\uJ k}^8/\Mpl^8}{\Mpl^2\left(1+c_6 \phi_{\uJ k}^2/\Mpl^2+c_7 \phi_{\uJ k}^4/\Mpl^4\right)^2},
}
where $c_i$ ($i=1, \cdots,7$) are constants given by
\beae{
c_1 &\coloneqq 24(\xi_g+\xi_\Gamma+3\xi_g^2+6\xi_g\xi_\Gamma),
\\
c_2 &\coloneqq 6(1+6\xi_g)(5\xi_g^2+13\xi_\Gamma^2+18\xi_g\xi_\Gamma+6\xi_g^3+24\xi_g^2\xi_\Gamma+24\xi_g\xi_\Gamma^2),
\\
c_3 &\coloneqq 6(1+6\xi_g)(\xi_g+\xi_\Gamma)(\xi_g^2+15\xi_\Gamma^2+12\xi_g\xi_\Gamma+6\xi_g^3+36\xi_g^2\xi_\Gamma+48\xi_g\xi_\Gamma^2),
\\
c_4 &\coloneqq 6(1+6\xi_g)\xi_\Gamma(\xi_g+\xi_\Gamma)^2(2\xi_g+7\xi_\Gamma+12\xi_g^2+30\xi_g\xi_\Gamma),
\\
c_5 &\coloneqq 6(\xi_g+\xi_\Gamma)^3(\xi_\Gamma+6\xi_g\xi_\Gamma)^2,
\\
c_6 &\coloneqq (1+6\xi_g)(\xi_g+2\xi_\Gamma),
\\
c_7 &\coloneqq \xi_\Gamma(1+6\xi_g)(\xi_g+\xi_\Gamma).
}
The cutoff scales in the large field limit and around the origin, corresponding to the inflation and reheating phases, take the following forms
\bae{
\Lambda_{\phi\to \infty}&\sim \frac{\Mpl}{\sqrt{6(\xi_\Gamma+\xi_g)}}\label{Eq. asymptotic large vev},
\\
\Lambda_{\phi\to0}&\sim \frac{\Mpl}{\sqrt{24(\xi_g+\xi_\Gamma+3\xi_g^2+6\xi_g\xi_\Gamma)}}\label{Eq. asymptotic origin}.
}
For the large value of $\xi_\Gamma$ or $\xi_g$, both cutoff scales become quite small compared to the Planck scale. Taking the limit either $\xi_\Gamma=0$ or $\xi_g=0$, we recover the same cutoff scales of the metric-Higgs or the Palatini-Higgs inflation, respectively.

\subsubsection{Unitarity violation}
In order to evaluate the \ac{EFT} validity, let us now compare the cutoff scales to physical energy scales. We expect that the perturbative unitarity is violated if the typical energy of excited particles exceeds the cutoff scales.

Let us first check what happens during inflation. On (quasi-)de Sitter background, the typical energy of excitations is the Hubble scale:
\bae{
E\sim H=\sqrt{\frac{U}{3\Mpl^2}}\simeq \sqrt{\frac{\lambda}{12}}\frac{\Mpl}{\xi_\Gamma+\xi_g}.
}
From Eq.~\eqref{Eq. asymptotic large vev}, the energy over the cutoff reads
\bae{
\eval{\frac{E}{\Lambda}}_{\phi\to\infty}\simeq\sqrt{\frac{\lambda}{2(\xi_\Gamma+\xi_g)}}.
}
This is much smaller than unity with a sufficiently large non-minimal coupling which is required for phenomenology. Therefore the unitarity problem would not be observed during inflation.

We then move on to the estimation in the reheating phase. Due to the self-interaction of the Higgs fields, the non-perturbative particle production, dubbed as preheating, takes place prior to the perturbative decay, in which the longitudinal modes with a large energy can be efficiently produced. As discussed in Refs.~\cite{Ema:2016dny,Ema:2021xhq}, the characteristic energy of the modes can be as large as the inflaton's potential energy at the end of inflation $U_{\mathrm{end}}$,
\bae{\label{eq: Eend}
E\sim U_{\mathrm{end}}^{1/4}\simeq \left(\frac{\lambda}{4}\right)^{1/4}\frac{\Mpl}{\sqrt{\xi_\Gamma+\xi_g}},
}
and it leads to
\bae{
\eval{\frac{E}{\Lambda}}_{\phi\to0}\simeq \lambda^{1/4}\sqrt{\frac{12(\xi_\Gamma+\xi_g+3\xi_g^2+6\xi_\Gamma\xi_g)}{\xi_\Gamma+\xi_g}}.
}
Given the constraint on the curvature perturbation~\eqref{Eq: curvature perturbation}, it seems that the ratio $E/\Lambda$ can be of order unity even when $\xi_g=0$ and $\lambda\sim 0.01$. This happens as this is a rough estimate, and Ref.~\cite{Rubio:2019ypq} showed that the unitarity is indeed preserved in the Palatini-Higgs limit. In this generalized case, however, there exists a monotonic increase of the ratio as $\xi_g$ gets larger. Therefore we expect that the energy of excited particles becomes generically greater than the \ac{UV} cutoff unless the self-coupling is tuned to be extremely small, and the perturbativity is lost during preheating phase, though a further investigation is needed for a rigorous statement. For precise inflationary predictions, our model would require \ac{UV}-extension.

\section{UV-extension}\label{Sec. UV-completion}
On theoretical grounds, the unitarity problem observed in the previous subsection is often recognized as a signal of new physics around the violation scale. Usually we expect that a new field (or multiple fields) with mass of order the \ac{UV} cutoff is excited at the scale and uniterizes the problematic low-energy theory. Therefore, we investigate the possibility of weakly-coupled \ac{UV}-completion of the hybrid metric-Palatini Higgs inflation by adding a new scalar field.

In the Einstein frame~\eqref{Eq. E-frame action}, the unitarity violation can be understood as a consequence of the curvature of the field-space $\calM_4$. Thus, we consider flattening the field-space as a way of UV-completion with help of an additional scalar field. In the language of geometry of the field-space, this is equivalent to embedding $\calM_4$ into a five-dimensional flat field-space $\mathbb{R}^5$. This geometrical procedure is successful in specifying a possible UV theory for the metric-Higgs inflation as investigated in Ref.~\cite{Mikura:2021clt}. In the following, we first review the embedding procedure and provide a possible candidate of a UV theory. We then give an implication that the minimum UV theory is healthy up to the Planck scale only in the limit of the metric-Higgs inflation. 

\subsection{Embedding approach and candidate for UV theory}
With use of the symmetry of the Higgs fields, the line element of the field-space $\calM_4$ can be rewritten in terms of the spherical coordinate as
\bae{\label{Eq. spherical exp.}
    \dd{s^2(\calM_4)}=G_{ij}\dd{\phi_{\uJ i}}\dd{\phi_{\uJ j}}\eqqcolon F^2(\rho_\uJ)\dd{\rho_\uJ^2}+\rho^2(\rho_\uJ)\dd{\Omega_3},
}
where $\rho_\uJ$ is the radial mode satisfying $\rho_\uJ^2=\phi_{\uJ i}^2$, $\dd\Omega_3$ is the angular line element of the three-dimensional sphere, and $F(\rho_\uJ)$ and $\rho(\rho_\uJ)$ are functions explicitly given by
\bae{
    F^2(\rho_\uJ)\coloneqq & \frac{1+(\xi_g+2\xi_\Gamma+6\xi_g^2+12\xi_g\xi_\Gamma)\frac{\rho_\uJ^2}{\Mpl^2}+(\xi_g\xi_\Gamma+\xi_\Gamma^2+6\xi_g^2\xi_\Gamma+6\xi_g\xi_\Gamma^2)\frac{\rho_\uJ^4}{\Mpl^4}}{\left(1+\xi_\Gamma\frac{\rho_\uJ^2}{\Mpl^2}\right)\left\{1+(\xi_g+\xi_\Gamma)\frac{\rho_\uJ^2}{\Mpl^2}\right\}^2},\label{Eq. def. F2} \\
    \rho^2(\rho_\uJ)\coloneqq &\frac{\rho_\uJ^2}{1+(\xi_g+\xi_\Gamma)\frac{\rho_\uJ^2}{\Mpl^2}}.\label{Eq. def. rho}
}
Once the spherical expression~\eqref{Eq. spherical exp.} is further deformed into
\bae{\label{Eq. embedding eq}
	\dd{s^2(\calM_4)}=\pqty{F^2(\rho_\uJ)-\pqty{
	\dv{\rho(\rho_\uJ)}{\rho_\uJ}}^2}\dd{\rho_\uJ^2}+\pqty{\dd{\rho(\rho_\uJ)}}^2+\rho^2(\rho_\uJ)\dd{\Omega_3},
}
the four-dimensional field-space $\calM_4$ can be viewed as a submanifold constrained on the hypersurface $z_\sur(\rho^2)$ in the five-dimensional flat field-space $\mathbb{R}^5$ as
\bae{\label{Eq. line element in 5D}
    \dd{s^2(\calM_4)}=\eval{\dd{s^2(\mathbb{R}^5)}}_{z=z_\sur(\rho^2)}=\eval{\dd{z^2}+\dd{\rho^2}+ \rho^2\dd{\Omega_3}}_{z=z_\sur(\rho^2)}.
}
Here $z_{\mathrm{sur}}$ is the hypersurface defined by\footnote{It should be noted that $F^2(\rho_\uJ)-\pqty{\rho^\prime(\rho_\uJ)}^2$ must be always non-negative, otherwise the new field $z$ has a wrong sign of the kinetic term and behaves as a harmful ghost field. In that case, two or more than two new fields would enable embedding to an Euclid space according to the Nash embedding theorem.}
\bae{\label{Eq. hypersurface definition}
    z_\sur(\rho^2)\coloneqq\int_0^{\rho_\uJ(\rho^2)}\sqrt{F^2(\tilde{\rho}_\uJ)-\left(\dv{\rho(\tilde{\rho}_\uJ)}{\tilde{\rho}_\uJ}\right)^2}\dd{\tilde{\rho}_\uJ}.
}

Physically, the hybrid metric-Palatini Higgs inflation model can be recognized as a low-energy \ac{EFT} obtained after integrating out a heavy mode in \ac{UV} theory and the unitarity issue can originate in a constraint being fixed on the hypersurface. In that sense, we expect that a healthy \ac{UV}-extension of our model can be constructed by introducing the hypersurface~\eqref{Eq. hypersurface definition} as an additional potential. One of the most simplest potential that acts as a multiplier is a following quartic coupling in the five-scalar's system:
\bae{\label{eq: additional potential}
    U_\mathrm{sur}(
    \rho^2,z)=\frac{\tilde{\lambda}}{4}\pqty{\rho^2-\rho^2_\sur(z)}^2,
}
where $\rho^2_\sur(z)$ denotes the inverse function of $z_\sur(\rho^2)$ and $\tilde{\lambda}$ is a coupling constant. With this hypersurface constraint, a possible candidate of \ac{UV} theory is then given by
\bae{\label{eq: S5}
	S_5=\int\dd[4]{x}\sqrt{-g}\left[\frac{\Mpl^2}{2}R-\frac{1}{2}\pqty{(\partial z)^2+(\partial\phi_i)^2}-\frac{\lambda}{4}(\phi_i^2)^2-U_\sur(\phi_i^2,z)\right],
}
where we adopt the Cartesian coordinate $\phi_i$ which satisfies $\rho^2=\phi_i^2$. As the gravitational operator is merely the Einstein--Hilbert term and the field-space is completely flat, the theory would be valid up to the Planck scale if the hypersurface potential is renormalizable or higher dimensional operators in $U_\sur$ are all suppressed by higher scales than $\Mpl$.

In the hybrid metric-Palatini Higgs inflation, the hypersurface potential $U_{\mathrm{sur}}$ cannot be described by a simple function due to the complexity of the hypersurface~\eqref{Eq. hypersurface definition}. Thus, in the following, we study the \ac{UV} cutoff scales at two distinct regions: during inflation and reheating. 
\subsection{During inflation}
Let us first consider the large $\rho_\uJ$ regime. In this approximation, Eq.~\eqref{Eq. hypersurface definition} can be analytically integrated as
\bae{
z_\sur\simeq \sqrt{\frac{1+6\xi_g}{\xi_g+\xi_\Gamma}}\log \rho_\uJ+\text{const}.
}
Combining this with Eq.~\eqref{Eq. def. rho}, an inverse function of the hypersurface reads
\bae{\label{Eq. inverse function during inflation}
\frac{\rho_\sur^2}{\Mpl^2}\simeq\frac{1}{\xi_g+\xi_\Gamma}\left(1-\frac{1}{Z^2(\xi_g+\xi_\Gamma)}\ee^{-2\sqrt{\frac{\xi_g+\xi_\Gamma}{1+6\xi_g}}\frac{z}{\Mpl}}\right),
}
where $Z$ is an integration constant. Thus the potential in the \ac{UV} theory is described by
\bae{
U_5&=\frac{\lambda}{4}(\phi_i^2)^2+U_\sur(\phi_i^2,z)\nonumber
\\
&\simeq \frac{\lambda}{4}(\phi_i^2)^2+\frac{\tilde{\lambda}}{4}\left[\phi_i^2-\frac{\Mpl^2}{\xi_g+\xi_\Gamma}\left(1-\frac{1}{Z^2(\xi_g+\xi_\Gamma)}\ee^{-2\sqrt{\frac{\xi_g+\xi_\Gamma}{1+6\xi_g}}\frac{z}{\Mpl}}\right)\right]^2.
}
We see that the flat potential can be realized along the $z$-direction. In the large $z$ limit, the $z$-dependence in the potential almost disappear due to an exponential suppression and the potential is simplified as
\bae{
U_5\simeq \frac{\lambda}{4}(\phi_i^2)^2+\frac{\tilde{\lambda}}{4}\left[\phi_i^2-\frac{\Mpl^2}{\xi_g+\xi_\Gamma}\right]^2.
}
Since this potential only contains renormalizable operators, the cutoff scale during inflation is uplifted to the Planck scale that stems from the Einstein--Hilbert term.

While the $z$ particle is almost massless during inflation, the radial mode $\rho$ is not whose mass $m_\rho$ can be read out from the above simplified potential as
\bae{
m_\rho^2=\partial_\rho^2U_5\simeq (3\lambda+2\tilde{\lambda})\frac{\Mpl^2}{\xi_g+\xi_\Gamma},
}
which corresponds to the cutoff during inflation~\eqref{Eq. asymptotic large vev}. Note that the $\rho$'s mass is much larger than the Hubble scale $H\sim \Mpl/(\xi_g+\xi_\Gamma)$. Thus the radial mode $\rho$ is hardly excited and hence the inflationary dynamics can be treated as a single-field one driven by $z$. Given the constraint~\eqref{Eq. inverse function during inflation}, the effective potential for $z$ is described by
\bae{
\eval{U_5}_{\rho_{\mathrm{sur}}}\simeq \frac{\lambda}{4}\rho_{\mathrm{sur}}^4\simeq \frac{\lambda\Mpl^4}{4(\xi_g+\xi_\Gamma)^2}\left(1-\frac{2}{Z^2(\xi_g+\xi_\Gamma)}\ee^{-2\sqrt{\frac{\xi_g+\xi_\Gamma}{1+6\xi_g}}\frac{z}{\Mpl}}\right),
}
which exhibits the same form as Eq.~\eqref{Eq. pheno potential}. Above discussion shows that our \ac{UV} theory can explain the origin of the low cutoff scale without changing the phenomenology.

\subsection{During reheating}
Let us move on to the small $\rho_\uJ$ region. To obtain an explicit form of the potential, we expand the integrand of Eq.~\eqref{Eq. hypersurface definition} around the origin and calculate $\rho^2_\sur(z)$. This can be done by the following Taylor expansion
\bae{
    \rho^2_\sur\simeq \eval{\left(\dv{z_\sur(\rho^2)}{\rho^2}\right)^{-1}}_{\rho=0}z+\frac{1}{2}\eval{\left(\dv{z_\sur(\rho^2)}{\rho^2}\right)^{-1}\dv{}{\rho^2}\left(\dv{z_\sur(\rho^2)}{\rho^2}\right)^{-1}}_{\rho=0}z^2+\cdots.
}
A resultant expression of the hypersurface constraint~\eqref{eq: additional potential} is given by
\bme{\label{Eq. potential around origin in five}
    U_\mathrm{sur}(r^2,z)\simeq \frac{\tilde{\lambda}}{4}\rho^4+\frac{\tilde{\lambda}}{2(\xi_g+3\xi_g^2+\xi_\Gamma+6\xi_g\xi_\Gamma)}z^2-\frac{g_{z^3}}{3!}z^3-\frac{g_{\rho^2z}}{2!}\rho^2z \\
    +\frac{\lambda_{z^4}}{4!}z^4+\frac{\lambda_{\rho^2z^2}}{2!2!}\rho^2z^2+\sum_{n=3}^\infty\left[\frac{1}{(n+2)!}\frac{z^{n+2}}{(\Lambda_{z^{n+2}})^{n-2}}+\frac{1}{2!n!}\frac{\rho^2 z^n}{(\Lambda_{\rho^2z^n})^{n-2}}\right],
}
where $g_i$, $\lambda_i$, and $\Lambda_i$ are functions of $\tilde{\lambda}$, $\xi_g$, and $\xi_\Gamma$. The subscript relates them with corresponding interactions, e.g., $g_{z^3}$ for the $z^3$-interaction.

The $z$ particle is no longer massless during reheating phase. It obtains the mass from the potential as
\bae{
m_z^2=\partial_z^2U_\mathrm{sur}\simeq\frac{\tilde{\lambda}}{\xi_g+3\xi_g^2+\xi_\Gamma+6\xi_g\xi_\Gamma},
}
which in turn explains the low cutoff of the low-energy theory~\eqref{Eq. asymptotic origin}. As the new scalar field $z$ is excited above the mass scale, this should be taken into account for the study of reheating.

\begin{figure}
    \centering
    \begin{tabular}{c}
        \begin{minipage}{0.5\hsize}
            \centering
            \includegraphics[width=0.95\hsize]{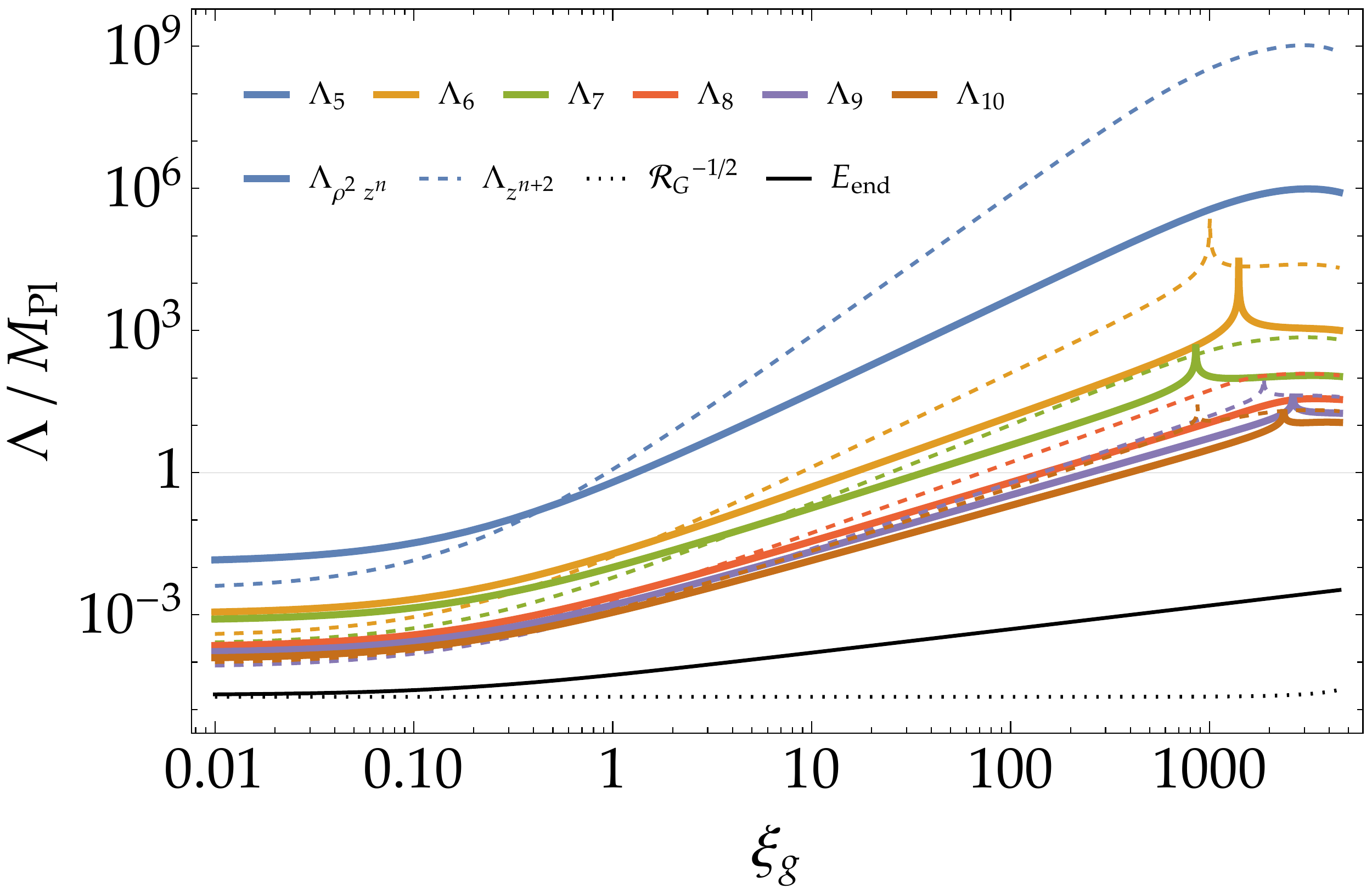}
        \end{minipage}
        \begin{minipage}{0.5\hsize}
            \centering
            \includegraphics[width=0.95\hsize]{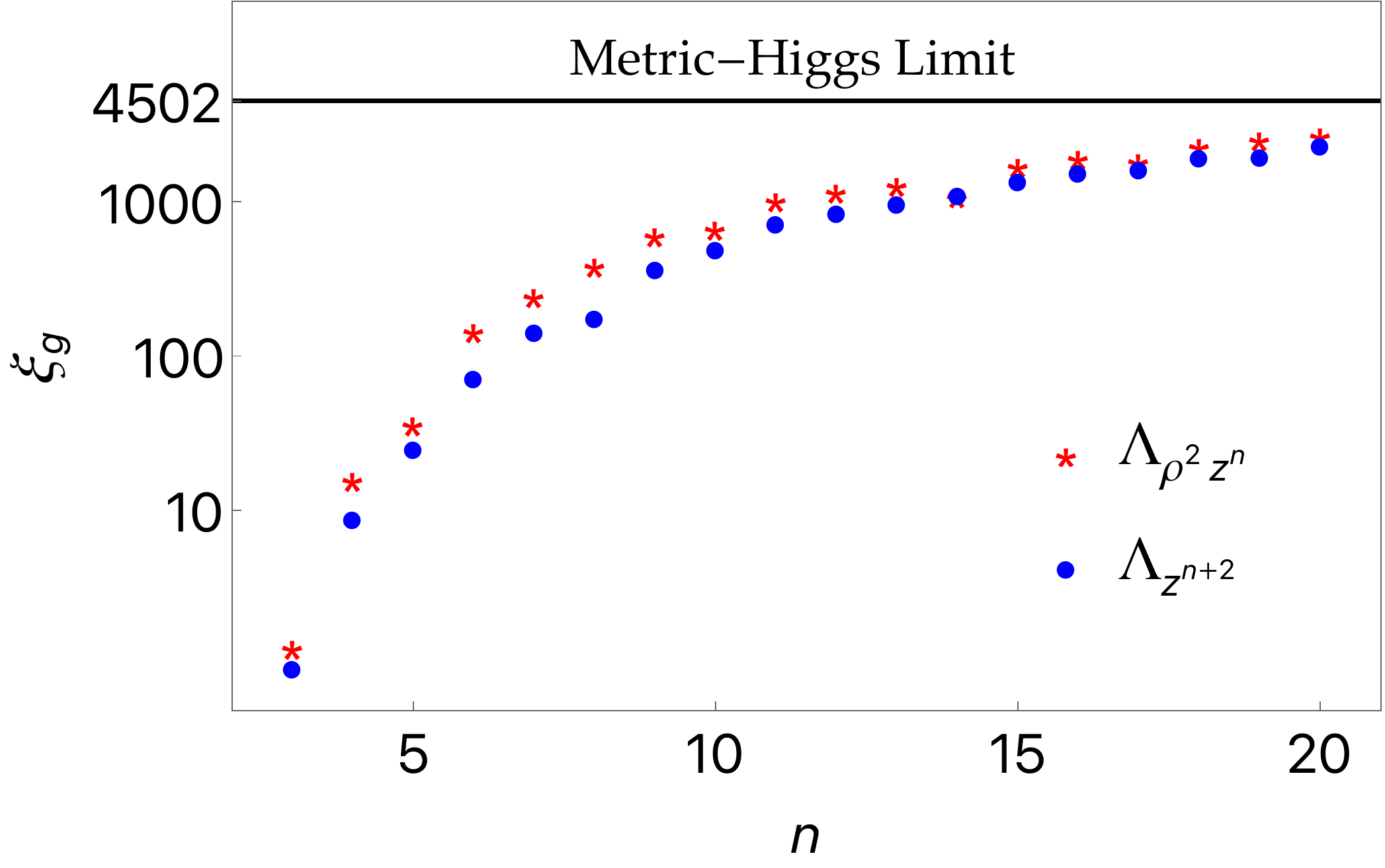}
        \end{minipage}
    \end{tabular}
    \caption{\emph{Left}: $\xi_g$-dependences of the cutoff scales where the remaining non-minimal coupling $\xi_\Gamma$ is fixed to satisfy the observed amplitude of the curvature perturbation~\eqref{Eq: curvature perturbation}. We employ $\lambda=\tilde{\lambda}=0.01$ and $N=55$. The rightest edge corresponds to the metric limit $\xi_\Gamma=0$.
    Colored solid and dashed lines correspond to the suppression scales of interactions $\rho^2 z^n $ and $ z^{n+2} $, respectively.
    The black dotted line indicates the inverse square-root of the field-space curvature $\calR_G$~\eqref{eq: RG}, i.e., the cutoff scale before the embedding.
    The black plane line is the energy scale at the end of inflation $E_\uend$~\eqref{eq: Eend}. The horizontal thin line represents the Planck scale $\Lambda=\Mpl$. 
    \emph{Right}: The minimum value of the non-minimal coupling $\xi_g$ satisfying $\Lambda\geq\Mpl$ with respect to the power of higher dimensional operators (i.e., the intersection of each line with the horizontal thin line). The black solid line is the limit of the metric-Higgs scenario with $\xi_\Gamma=0$.
    }
    \label{fig: required coupling vs dimension}
\end{figure}

Let us now see the cutoff scale of the five-scalar theory. It can be determined from higher-dimensional operators represented by the last two collective terms in Eq.~\eqref{Eq. potential around origin in five} through the power-counting approach. The minimum of their suppression scales i.e., $\Lambda_{\rho^2z^3}$, $\Lambda_{z^5}$, $\cdots$, corresponds to the cutoff of the theory. The $\xi_g$-dependences of some higher dimensional operators are depicted in the left panel of Fig.~\ref{fig: required coupling vs dimension}, where $\xi_\Gamma$ is fixed by the \ac{CMB} normalization~\eqref{Eq: curvature perturbation}.
One sees that, while the cutoff scale estimated by the field-space curvature before embedding is lower than the dynamical scale during the preheating, the embedding successfully uplifts all cutoff scales higher than the dynamical scale for any $\xi_g$. However, it also implies that the UV-completion up to the Planck scale can be accomplished only in the metric limit $\xi_\Gamma\to0$ if sufficiently higher dimensional operators are considered. This behavior is clear in the right panel of Fig.~\ref{fig: required coupling vs dimension}, which shows the minimum value of $\xi_g$ to satisfy $\Lambda\geq\Mpl$ for each operator. It implies that $\xi_g$ eventually asymptotes to the metric-Higgs limit $\xi_\Gamma=0$ for a sufficiently large $n$. Expecting from the behavior, UV-completion with one additional scalar field might be realized only for the metric-Higgs inflation.

\section{Generalizations}\label{Sec. Generalizatons}

The hybrid metric-Palatini Higgs inflation~\eqref{Eq. action for the hybrid metric-Palatini} is a minimal theory that bridges between the metric-Higgs and Palatini-Higgs inflation models. However, the symmetry of the theory still allows other operators up to mass-dimension four. In this section, we study effects with each of the $\Mg^2R_\uJ[g]$, $R_\uJ^2[g]$, and $R_\uJ^2[\Gamma]$ terms. While we may consider a curvature mixing $R_\uJ[g]R_\uJ[\Gamma]$ from the view point of EFT, a mixing of the metric and Palatini curvature terms is not allowed because it is shown in Ref.~\cite{Capozziello:2015lza} that unhealthy modes inevitably arise in the presence of such coupling. We may also include the sum $R_\uJ^2[g] + R_\uJ^2[\Gamma]$ term on top of our hybrid model. However, it gives another class of inflation with two additional scalar fields and the inflationary dynamics can be altered significantly, so it’s beyond the scope of the paper. We will investigate this case in our future work.

\subsection{$\Mg^2R_\uJ[g]$ term}\label{Additional EH}
From an \ac{EFT} perspective, we may add the metric curvature $R[g]$ in a linear form with a gauge coupling $M_g$ in addition to the original Lagrangian~\eqref{Eq. action for the hybrid metric-Palatini}. The Lagrangian is defined by
\bae{\label{eq-EH-Jordan}
\frac{\calL_R}{\sqrt{-g_{\uJ}}}=\frac{1}{2}\left(\MG^2+\xi_\Gamma\phi_{\uJ i}^2\right)R_\uJ[\Gamma]+\frac{1}{2}\left(\Mg^2+\xi_g \phi_{\uJ i}^2\right)R_\uJ[g]-\frac{1}{2}g^{\mu\nu}_{\uJ}\partial_\mu\phi_{\uJ i}\partial_\nu\phi_{\uJ i}-\frac{\lambda}{4}(\phi_{\uJ i}^2)^2,
}
where a scale of the Einstein--Hilbert term formed by the Palatini curvature is set to be $M_\Gamma$ for generality. We assume $\MG^2+\Mg^2=\Mpl^2$ in order to recover the usual Einstein--Hibert term around the electroweak vacuum $\phi_\uJ\sim0$. This is the most general up to dimension four including the linear curvature and the hybrid metric-Palatini Higgs inflation can be obtained by simply taking the limit $\Mg=0$.

To explore the role of the added Einstein--Hilbert term, let us move to the Einstein frame. The Euler--Lagrange equation for the connection relates the metric and Palatini curvatures as shown in Sec.~\ref{Hybrid metric-Palatini Higgs inflation}, yielding the corresponding metric theory as
\bae{
\frac{\calL_R}{\sqrt{-g_{\uJ}}}=\frac{\Mpl^2}{2}\left[1+\left( \xi_{\Gamma}+\xi_g\right)\frac{\phi_{\uJ i}^2}{\Mpl^2} \right] R_\uJ-\frac{1}{2}\left(\delta_{ij} -\frac{6\xi_\Gamma^2\frac{\phi_{\uJ i}\phi_{\uJ j}}{M^2_\Gamma}}{1+\xi_\Gamma\frac{\phi_{\uJ i}^2}{M^2_\Gamma}} \right) g^{\mu\nu}_\uJ\partial_{\mu} \phi_{\uJ i} \partial_\nu \phi_{\uJ j}-\frac{\lambda}{4}\left(\phi_{\uJ i}^2\right)^2.
}
Performing the Weyl transformation
\bae{
    g_{\mu\nu} = \left[ 1+\left( \xi_{\Gamma} + \xi_g \right) \frac{\phi_{\rm J}^2}{M_{\rm pl}^2} \right] g_{{\rm J}\mu\nu} \eqqcolon \Omega^2 g_{{\rm J}\mu\nu},
}
the Lagrangian in the Einstein frame can be obtained as (the superscript ``EH'' stands for the Einstein--Hilbert correction)
\bae{
    \frac{\mathcal{L}_{R}}{\sqrt{-g}} =\frac{\Mpl^2}{2} R-\frac{1}{2}G^{\mathrm{EH}}_{ij}\partial_{\mu} \phi_{{\rm J}i} \partial^{\mu} \phi_{{\rm J}j} - \frac{\lambda}{4\Omega^4}(\phi_{\uJ i}^2)^2,
}
with the field-space metric
\bae{
  G^{\mathrm{EH}}_{ij}\coloneqq \frac{1}{\left[ 1+ \left(\xi_{\Gamma} + \xi_g\right) \frac{\phi_{\uJ i}^2}{\Mpl^2} \right]^2\left(\frac{M_{\Gamma}^2}{\Mpl^2} +\xi_{\Gamma} \frac{\phi_{\uJ i}^2}{\Mpl^2}\right)}\left[ \left( 1+(\xi_{\Gamma} +\xi_g) \frac{\phi_{\uJ i}^2}{\Mpl^2}  \right) \left(\frac{M_{\Gamma}^2}{\Mpl^2} +\xi_{\Gamma}\frac{\phi_{\uJ i}^2}{\Mpl^2}\right) \delta_{ij} \right. \nonumber \\
  \left. +6 \left( -\xi_{\Gamma}^2 +\left(\xi_{\Gamma} +\xi_g \right)^2 \frac{M_{\Gamma}^2}{\Mpl^2} + \xi_{\Gamma} \xi_g \left( \xi_{\Gamma} +\xi_g \right) \frac{\phi_{\uJ i}^2}{\Mpl^2} \right) \frac{\phi_{\uJ i} \phi_{\uJ j}}{\Mpl^2} \right].
    }
Given the field-space metric, the cutoff scales during inflation and the reheating can be estimated geometrically from the Ricci curvature
\bae{
    \calR_G^{\mathrm{EH}}\Mpl^2=\frac{d_1+d_2x^2+d_3x^4+d_4x^6+d_5x^8}{\qty(f^2+d_6x^2+d_7x^4)^2},
}
where $x^2=\phi_{\uJ k}^2/\Mpl^2$, $f=M_\Gamma/\Mpl$, and coefficients $d_i$ ($i=1,\cdots, 7$) are explicitly given by
\beae{
    d_1&\coloneqq-72 f^2\xi_\Gamma^2 + 24 f^4 (\xi_g + \xi_\Gamma) (1 + 3\xi_g + 3\xi_\Gamma), \\
    d_2&\coloneqq \bmte{6 \left[6\xi_\Gamma^3 (-1 + 6\xi_\Gamma) + f^4 (\xi_g + \xi_\Gamma)^2 (1 + 6\xi_g + 6\xi_\Gamma) (5 + 6 \xi_g + 6\xi_\Gamma)\right. \\ 
    \left.- 2 f^2\xi_\Gamma (\xi_g + \xi_\Gamma) \qty(-4 + 12 \xi_g (-1 + 3 \xi_\Gamma) + 3 \xi_\Gamma (5 + 12 \xi_\Gamma))\right],} \\
    d_3&\coloneqq\bmte{ 6 (\xi_g + \xi_\Gamma) \left[-2 (2 + 6 \xi_g - 3 \xi_\Gamma) \xi_\Gamma^2 (-1 + 6 \xi_\Gamma) + f^4 (\xi_g + \xi_\Gamma)^2 (1 + 6 \xi_g + 6 \xi_\Gamma)^2 \right.\\ 
    \left.+ 2 f^2 \xi_\Gamma (\xi_g + \xi_\Gamma) \qty(5 + 36 \xi_g (1 + \xi_g) + 9 (1 - 4 \xi_\Gamma) \xi_\Gamma)\right],} \\
    d_4&\coloneqq 6 (1 + 6 \xi_g) \xi_\Gamma (\xi_g + \xi_\Gamma)^2 \bqty{(5 + 6 \xi_g - 12 \xi_\Gamma) \xi_\Gamma + 2 f^2 (\xi_g + \xi_\Gamma) (1 + 6 \xi_g + 6 \xi_\Gamma)}, \\
    d_5&\coloneqq6 (\xi_g + \xi_\Gamma)^3 (\xi_\Gamma + 6 \xi_g \xi_\Gamma)^2, \\
    d_6&\coloneqq\xi_\Gamma - 6 \xi_\Gamma^2 + f^2 (\xi_g + \xi_\Gamma) (1 + 6 \xi_g + 6 \xi_\Gamma), \\
    d_7&\coloneqq\xi_\Gamma(1 + 6 \xi_g) (\xi_g + \xi_\Gamma).
}

Expecting that inflation occurs at the large field value, non-minimal coupling terms, $\xi_\Gamma\phi_{\uJ i}^2 R_\uJ[\Gamma]$ and $\xi_g \phi_{\uJ i}^2 R_\uJ[g]$ in the Jordan frame, play important roles than the Einstein--Hilbert terms, and hence inflationary phenomenology would not be changed by the new operator. Indeed, in the large field limit, the radial eigenvalue coincides with $F^2(\phi_\uJ)$ in Eq.~\eqref{eq: K(phiJ) in the large field limit}. Correspondingly, the cutoff scale in the large-field limit $x\gg1$ is free from the ratio of the two gauge couplings $f$ as
\bae{
    \calR_G^{\mathrm{EH}}\underset{x\gg1}{\sim}\frac{6(\xi_g+\xi_\Gamma)}{\Mpl^2},
}
which is equivalent to the case of the hybrid metric-Palatini Higgs inflation. 

On the other hand, the dynamics around the origin $\phi_{\uJ i}=0$ is sensitive to the new term and the curvature around the origin is significantly altered as
\bae{
    \calR_G^{\mathrm{EH}}\underset{x\sim0}{\sim}-\frac{72\xi_\Gamma^2}{M_\Gamma^2}+\frac{24(\xi_g+\xi_\Gamma)(1+3\xi_g+3\xi_\Gamma)}{\Mpl^2}.
}
For $M_\Gamma=\Mpl$, the $\xi_\Gamma^2$ terms are cancelled and that is the reason why the cutoff scales are proportional to $1/\sqrt{\xi_\Gamma}$ in the Palatini-Higgs inflation. However, in a general case with $M_\Gamma<\Mpl$, the first term dominates and the cutoff is significantly lowered as $\Lambda_{\mathrm{EH}}\sim\abs{\calR_G^{\mathrm{EH}}}^{-1/2}\sim M_\Gamma/\xi_\Gamma$. The negativity of the field-space curvature may imply that the embedding approach fails. In order to embed the curved field-space in a five-dimensional space successfully, it is necessary to satisfy $F^2(\rho_\uJ)-\qty(\rho^\prime(\rho_\uJ))^2\geq0$ as introduced in Eq.~\eqref{Eq. embedding eq}.
In this generalization, as functions $F^2$ and $\rho^2(\rho_\uJ)$ are given by
\beae{
    F^2(\rho_\uJ)&\coloneqq\frac{x^2 \xi_\Gamma \bqty{1 - 6 \xi_\Gamma + x^2 (1 + 6 \xi_g) (\xi_g + \xi_\Gamma)} + f^2 \bqty{1 + x^2 (\xi_g + 6 \xi_g^2 + \xi_\Gamma + 12 \xi_g \xi_\Gamma + 6 \xi_\Gamma^2)}}{(f^2 + x^2 \xi_\Gamma) \bqty{1 + x^2 (\xi_g + \xi_\Gamma)}^2}, 
    \\
    \rho^2(\rho_\uJ)&\coloneqq\frac{\rho_\uJ^2}{1+x^2(\xi_g+\xi_\Gamma)},
}
the condition cannot be kept for $f<\sqrt{3\xi_{\Gamma}/(1+3\xi_{\Gamma})} $ which is of order unity for large $ \xi_{\Gamma} $ as can be seen from
\bae{
    F^2(\rho_\uJ)-\pqty{\rho^\prime(\rho_\uJ)}^2\underset{x\sim0}{\sim}\frac{2\xi_\Gamma(-3\xi_\Gamma+f^2(1+3\xi_\Gamma))}{f^2}x^2.
}
It implies that $\Mg^2 R[g]$ must not exist. In other words, $\MG$ must to be equal to $\Mpl$, to construct a healthy UV theory with five scalar fields.
\subsection{$R^2_\uJ[g]$ term}\label{R2 section}
Let us consider an extension with a gravitational operator that is quadratic in the metric curvature. The Lagrangian of our interest is given by
\bae{\label{eq-R2-Jordan}
\frac{\calL_{R^2[g]}}{\sqrt{-g_{\uJ}}}=\frac{\Mpl^2}{2}\left(1+\xi_\Gamma\frac{\phi_{\uJ i}^2}{\Mpl^2}\right)R_\uJ[\Gamma]+\frac{1}{2}\xi_g \phi_{\uJ i}^2R_\uJ[g]+\frac{\alpha}{4} R^2_\uJ[g]-\frac{1}{2}g^{\mu\nu}_{\uJ}\partial_\mu\phi_{\uJ i}\partial_\nu\phi_{\uJ i}-\frac{\lambda}{4}(\phi_{\uJ i}^2)^2,
}
where $\alpha$ is a dimensionless parameter. This extension is particularly interesting because, in the metric-Higgs inflation, the additional quadratic term makes one scalar mode in the metric dynamical and it acts as a UV-completing field. Thus, in the following, we investigate if such term can also unitarize the hybrid metric-Palatini Higgs inflation.

In order to investigate the UV aspect, let us first derive the Einstein-frame expression of Eq.~\eqref{eq-R2-Jordan}. With the Euler--Lagrange equation for the connection, the action~\eqref{eq-R2-Jordan} reduces to 
\bme{
    \frac{\mathcal{L}_{R^2[g]}}{\sqrt{-g_{\uJ}}}=\frac{\Mpl^2}{2}\left[1+\left( \xi_{\Gamma}+\xi_g\right)\frac{\phi_{\uJ i}^2}{\Mpl^2} \right] R_\uJ+\frac{\alpha}{4} R_\uJ^2 \\
    -\frac{1}{2}\left(\delta_{ij} -\frac{6\xi_\Gamma^2\frac{\phi_{\uJ i}\phi_{\uJ j}}{\Mpl^2}}{1+\xi_\Gamma\frac{\phi_{\uJ i}^2}{\Mpl^2}}\right)g^{\mu\nu}_\uJ \partial_{\mu} \phi_{\uJ i} \partial_\nu \phi_{\uJ j}-\frac{\lambda}{4}\left(\phi_{\uJ i}^2\right)^2.
}
Now that this is nothing but a well-known $f(R)$ theory, the Einstein-frame expression can be easily obtained. By defining a scalar field $\psi$ (often called scalaron) through
\bae{
\sqrt{\frac{2}{3}}\frac{\psi}{\Mpl}\coloneqq \log\left[1+\left( \xi_{\Gamma}+\xi_g \right)\frac{\phi_{\uJ i}^2}{\Mpl^2}+\alpha \frac{R_\uJ}{\Mpl^2}\right],
}
and performing the Weyl transformation
\bae{\label{Eq. E-frame with R2 trans}
 g_{\mu\nu} \coloneqq \ee^{\sqrt{\frac{2}{3}}\frac{\psi}{\Mpl}} g_{\uJ\mu\nu},
}
we obtain the Einstein-frame Lagrangian as
\bae{\label{Eq. E-frame with R2}
\frac{\mathcal{L}_{R^2[g]}}{\sqrt{-g}}=\frac{\Mpl^2}{2}R-\frac{1}{2}g^{\mu\nu}\partial_\mu\psi\partial_\nu\psi-\frac{1}{2}\ee^{-\sqrt{\frac{2}{3}}\frac{\psi}{\Mpl}}\left(\delta_{ij}-\frac{6\xi_\Gamma^2\frac{\phi_{\uJ i}\phi_{\uJ j}}{\Mpl^2}}{1+\xi_\Gamma\frac{\phi_{\uJ i}^2}{\Mpl^2}}\right)g^{\mu\nu}\partial_\mu\phi_{\uJ i}\partial_\nu\phi_{\uJ j}-U_{R^2},
}
where $U_{R^2[g]}$ denotes the following potential:
\bae{
U_{R^2[g]}\coloneqq \frac{\lambda}{4}\ee^{-2\sqrt{\frac{2}{3}}\frac{\psi}{\Mpl}}(\phi_{\uJ i}^2)^2+\frac{\Mpl^4}{4\alpha}\left[1-\ee^{-\sqrt{\frac{2}{3}}\frac{\psi}{\Mpl}}\left(1+(\xi_\Gamma+\xi_g)\frac{\phi_{\uJ i}^2}{\Mpl^2}\right)\right]^2.
}

It can be checked that the Einstein-frame expression with $\xi_\Gamma=0$ limit coincides with the mixed Higgs-$R^2$ model defined in the Riemannian geometry~\cite{He:2018gyf}. In this limit, while the field-space is still curved, the unitarity problem is not observed as the field-space curvature is uplifted to the Planck scale and higher dimensional operators in the potential are also suppressed by higher scales than $\Mpl$ if the coupling constant $\alpha$ is of order $\calO(\xi^2)$.\footnote{In the case of the metric-Higgs inflation, this has been confirmed in Ref.~\cite{Ema:2019fdd}.} It can be understood that UV-completion of the metric-Higgs inflation is realized by a scalar degree of freedom arose from the $R^2$ operator.\footnote{For a detail, see e.g. Refs.~\cite{Salvio:2014soa,Kannike:2015apa,Salvio:2016vxi,Salvio:2017qkx,Ema:2019fdd,Ema:2020zvg,Ema:2020evi}.} 

The only difference generated by the $\xi_\Gamma$ is the derivative coupling among the Higgs fields, which implies the existence of a strong curvature even in the five dimensional field-space. To certify this, it is enough to calculate the field-space curvature around the origin. The curvature depends explicitly on the non-minimal coupling $\xi_\Gamma$ as
\bae{
\abs{\calR_G\Mpl^2}=\frac{2}{3}(5+108\xi_\Gamma^2),
}
so that the cutoff scale of the theory~\eqref{Eq. E-frame with R2} is still low as
\bae{
\Lambda_{R^2[g]}=\sqrt{\frac{3}{2(5+108\xi_\Gamma^2)}}\Mpl\underset{\xi_\Gamma\gg1}{\sim}\frac{\Mpl}{10\xi_\Gamma}.
}
This clearly indicates that the scalaron coming from $R^2[g]$ operator does not play a role of a UV-completing field in the hybrid metric-Palatini Higgs inflation.

\subsection{$R^2_\uJ[\Gamma]$ term}\label{R2Gamma section}
Let us turn to the case with the $R_\uJ^2[\Gamma]$ term on top of the hybrid metric-Palatini Higgs inflation. The Lagrangian is given by
\bae{\label{Eq. R2Gamma Lagrangian}
    \frac{\calL_{R^2[\Gamma]}}{\sqrt{-g_\uJ}}=\frac{\Mpl^2}{2}\left(1+\xi_\Gamma \frac{\phi_{\uJ i}^2}{\Mpl^2}\right)R_\uJ[\Gamma]+\frac{1}{2}\xi_g\phi_{\uJ i}^2R_\uJ[g]+\frac{\beta}{4}R_\uJ^2[\Gamma]-\frac{1}{2}\partial_\mu\phi_{\uJ i}\partial^\mu\phi_{\uJ i}-\frac{\lambda}{4}(\phi_{\uJ i}^2)^2,
}
where $\beta$ is a dimensionless parameter characterizing the additional operator. Following the same approach used in the previous subsection, we define a scalar field $\psi$ by
\bae{
\frac{\psi}{\Mpl}\coloneqq 1+\xi_\Gamma\frac{\phi_{\uJ i}^2}{\Mpl^2}+\beta \frac{R_\uJ[\Gamma]}{\Mpl^2}.
}
It gives the following equivalent scalar-tensor theory in the Jordan frame:
\bae{
    \frac{\mathcal{L}_{R^2[\Gamma]}}{\sqrt{-g_{\uJ}}}=\frac{1}{2}\xi_g\phi_{\uJ i}^2 R_\uJ[g]+\frac{1}{2}\psi\Mpl R_\uJ[\Gamma]-\frac{1}{2}g^{\mu\nu}_\uJ \partial_{\mu} \phi_{\uJ i} \partial_\nu \phi_{\uJ i}-U_{R^2[\Gamma]},
}
where the potential $U_{R^2[\Gamma]}$ contains both the Higgses' quartic coupling and the contribution from the quadratic curvature as
\bae{
U_{R^2[\Gamma]}=\frac{\lambda}{4}(\phi_{\uJ i}^2)^2+\frac{\Mpl^4}{4\beta}\left(\frac{\psi}{\Mpl}-1-\xi_\Gamma\frac{\phi_{\uJ i}^2}{\Mpl}\right)^2.
}
Since the Lagrangian is linear in the Palatini curvature, one can solve the constraint equation for the connection with Eq.~\eqref{Eq. EoM for connection} and the theory reduces to
\bae{
    \frac{\calL_{R^2[\Gamma]}}{\sqrt{-g_\uJ}}=\frac{\Mpl^2}{2}\left(\frac{\psi}{\Mpl}+\xi_g\frac{\phi_{\uJ i}^2}{\Mpl^2}\right)R_\uJ [g]+\frac{3\Mpl}{4\psi}(\partial\psi)^2-\frac{1}{2}\partial_\mu\phi_{\uJ i}\partial^\mu\phi_{\uJ i}-U_{R^2[\Gamma]}.
}
Here it should be noted that the scalar field $\psi$ is not dynamical in the Palatini limit ($\xi_g=0$) due to a certain relation between the non-minimal coupling and $\psi$'s kinetic term~\cite{Olmo:2011uz}. However the metric curvature coupled to the Higgs fields breaks the relation in our hybrid case and there exists an additional dynamical component in the theory.\footnote{See Ref.~\cite{Tamanini:2013ltp} for the details.}

By rescaling the metric through
\bae{
g_{\mu\nu}\coloneqq\Omega^2g_{\uJ \mu\nu}=\left(\frac{\psi}{\Mpl}+\xi_g\frac{\phi_{\uJ i}^2}{\Mpl^2}\right)g_{\uJ \mu\nu},
}
one can arrive at the expression in the Einstein frame as
\bae{
 \frac{\calL_{R^2[\Gamma]}}{\sqrt{-g}}=\frac{\Mpl^2}{2}R[g]+X(\phi_{\uJ i}, \psi)-\frac{1}{\Omega^4}U_{R^2[\Gamma]},
}
where $X$ denotes the kinetic term with five scalars given explicitly by
\bae{\label{Eq. conical kinetic term}
X=-\frac{1}{2}\left[\frac{-3\Mpl \Omega^2+3\psi}{2\Omega^4 \psi}(\partial \psi)^2+\frac{6\xi_g}{\Omega^4}\frac{\phi_{\uJ i}}{\Mpl}\partial_\mu\phi_{\uJ i}\partial^\mu \psi+\left(\frac{\delta_{ij}}{\Omega^2}+\frac{6\xi_g^2}{\Omega^4}\frac{\phi_{\uJ i}\phi_{\uJ j}}{\Mpl^2}\right)\partial_\mu\phi_{\uJ i}\partial^\mu\phi_{\uJ j}\right].
}
One can calculate the Ricci scalar of the above five-dimensional field-space as
\bae{
\calR_G\Mpl^2 = 6\xi_g-\frac{4}{3}+\frac{(2+6\xi_g)\Mpl}{\xi_g }\frac{\psi}{\phi_{\uJ i}^2}.
}
It is easy to see that the curvature diverges at $ \phi_{\uJ i} =0 $ except for $ \xi_g =-1/3 $, which indicates a conical singularity at this point. In this case, we cannot determine the UV cutoff of this model at $ \phi_{\uJ i} =0 $ in a simple way. A further study is needed while it is beyond the scope of this paper. 

\section{Conclusions}\label{Sec. Conclusions}
In this paper, we have proposed the hybrid metric-Palatini Higgs inflation that is an extension of the Higgs inflation to the hybrid metric-Palatini gravity where we introduce non-minimal couplings between Higgs and both the metric-type and the Palatini-type Ricci curvatures. In section~\ref{Hybrid metric-Palatini Higgs inflation}, we have first studied the inflationary phenomenology focusing on the radial mode of the Higgs fields. We have found that the spectral index of the curvature perturbation is independent of non-minimal coupling parameters and exhibits an attractor behavior converging to $\ns\sim 0.964$ which is perfectly consistent with current \ac{CMB} observations. On the other hand, the tensor-to-scalar ratio is sensitive to those parameters where the maximum value is $r\sim 10^{-3}$ in the metric-Higgs limit and it can be much smaller as the model gets closer to the Palatini-Higgs limit. In section~\ref{Sec. UV-completion} we have investigated the theoretical consistency of the hybrid metric-Palatini Higgs inflation. As it can be understood that the unitarity violation of our model originates in a strong curvature of the filed-space spanned by the Higgs fields, we have investigated one possible example of \ac{UV}-extension by demanding that a new scalar field completely eliminates the Ricci curvature of the field-space. We then found that cutoff scales of the UV theory would be generically greater than the dynamical scale as depicted in the left panel of Fig.~\ref{fig: required coupling vs dimension}, ensuring that the UV theory can be viewed as a weakly-coupled \ac{EFT}. However, we expect that the Planck-scale \ac{EFT} can be only realized in the limit of the metric-Higgs inflation scenario because, as can be seen in the right panel of Fig.~\ref{fig: required coupling vs dimension}, a required value of $\xi_g$ increases as one sees operators with mass-dimension larger than four and it asymptotes to the metric-Higgs limit $\xi_\Gamma=0$. Possible generalizations of the hybrid metric-Palatini Higgs inflation are studied in section~\ref{Sec. Generalizatons} with concrete examples: $M_g^2R[g]$, $R^2[g]$, or $R^2[\Gamma]$ term. For the first generalization, we found that the additional Einstein--Hilbert term cannot exist to have a healthy UV theory with one additional scalar. The second possibility has been attracting some attentions in that it generates a dynamical scalar field that unitarizes the metric-Higgs inflation. However, in our case, the scalar field coming from the $R^2[g]$ operator can not play a role of UV-completing field. For the generalization with the $R^2[\Gamma]$ term, we found the resulting field-space in five dimension has a conical singularity and the new scalar field has no dynamical role for UV completion.

Satisfactory understanding of UV-completion of metric/Palatini-Higgs and our hybrid metric-Palatini Higgs inflation is yet to be obtained and further investigations are required. For instance, one can extend the potential~\eqref{eq: additional potential} to more general form (e.g. exponential) to cure the higher-order operators such that the UV-completion up to Planck scale is realized. Although the affine connection is determined by the Euler--Lagrange constraint in Palatini-Higgs and our hybrid model, it might be a dynamical degree at high-energy regime. Thus, it would be interesting to scrutinize effects of additional dynamical components in the connection. Generalizations to this direction, Einstein--Cartan or metric-affine for instance, may give us insights on the origin of the inevitably small UV-cutoff.

\acknowledgments

We are grateful to Yohei Ema and Kyohei Mukaida for helpful discussions. This work is supported by JSPS KAKENHI Grants No.~JP22J22254 (Y.M.) and 
No.~JP21K13918 (Y.T.).

\bibliographystyle{JHEP}
\bibliography{Bib}
\end{document}